\documentclass[3p,times,twocolumn]{elsarticle}
\usepackage{amssymb,amsmath,multirow,dcolumn,bm,float,graphicx,epstopdf}
\usepackage{calrsfs}  
\usepackage{epsfig}
\usepackage{soul}
\makeatletter

\usepackage{amssymb}
\usepackage{graphicx,color}
\usepackage[usenames,dvipsnames]{xcolor}
\usepackage{subfigure}
\usepackage{booktabs}

\colorlet{darkgreen}{green!50!black}
\colorlet{brightyellow}{yellow!75!red}%
\colorlet{orange}{red!50!yellow}
\colorlet{darkblue}{blue!60!black}
\colorlet{darkred}{red!80!black}

\usepackage{soul}

\usepackage{mathtools}
\usepackage{mathrsfs}
\usepackage{bm}
\usepackage{relsize}
\usepackage{indentfirst}

\usepackage{hyperref}
\hypersetup{
    unicode=false,          
    pdftoolbar=true,        
    pdfmenubar=true,        
    pdffitwindow=false,     
    pdfstartview={FitH},    
    pdfnewwindow=true,      
    colorlinks=true,       
    linkcolor=red,          
    citecolor=red,        
    filecolor=cyan,         
    urlcolor=magenta        
}

\newcommand{\nc}{\newcommand}
\nc{\lamnr}{\lambda_{nr}}
\nc{\lamr}{\lambda_{r}}
\nc{\lamk}{\lambda_{k}}
\nc{\gp}{g({\bp})}
\nc{\gpp}{g({\bp'})}
\nc{\gpz}{g({\bp''})}
\nc{\gps}{g^{*}({\bp})}
\nc{\gpzs}{g^{*}({\bp''})}
\nc{\gpps}{g^{*}({\bp'})}
\nc{\bxi}{{\bf \xi}}
\nc{\bp}{{\bf p}}
\nc{\bpp}{{\bf p'}}
\nc{\bpz}{{\bf p''}}
\nc{\bk}{{\bf k}}
\nc{\bkp}{{\bf k'}}
\nc{\bkz}{{\bf k''}}
\nc{\bPi}{{\bf \Pi}}
\nc{\bera}{\langle}
\nc{\ket}{\rangle}
\nc{\bq}{{\bf q}}
\nc{\bqp}{{\bf q'}}
\nc{\tpi}{\tilde{\pi}}
\nc{\bpi}{\boldsymbol \pi}
\nc{\btpi}{\tilde{\boldsymbol \pi}}
\nc{\andre}[1]{\textcolor{red}{#1}}

\journal{Physics Letters B}

\begin{document}
\begin{frontmatter}
\title{ Proton quark distributions from a light-front Faddeev-Bethe-Salpeter approach }

\author[1]{E.~Ydrefors}
\ead{ydrefors@kth.se}
\address[1]{Institute of Modern Physics, Chinese Academy of Sciences, Lanzhou 730000, China}
\author[2]{T.~Frederico}
\ead{tobias@ita.br}
\address[2]{Instituto Tecnol\'ogico de Aeron\'autica,  DCTA, 12228-900 S\~ao Jos\'e dos Campos,~Brazil}

\date{\today}

\begin{abstract}
The projection onto the Light-Front  of a Minkowski space Faddeev-Bethe-Salpeter equation model truncated at the  valence level is applied to  study the proton structure with constituent quarks.
The dynamics of the model has  built-in: (i) a bound diquark brought by a contact interaction, and (ii) the separation by $\sim \Lambda_{QCD}$ of the infrared and ultraviolet interaction regions.
The model parameters are  fine tuned to reproduce the proton Dirac electromagnetic form factor and mass. From that, the non-polarized longitudinal  and transverse momentum distributions  were computed. The results for the evolved non-polarized valence parton distributions suggest that: (i) the explicit consideration of the spin degree of freedom of both quark and diquark seems not relevant to it, and (ii) the comparison with the global fit from  the NNPDF4.0 calls for higher Fock components of the wave function beyond the valence one.

\end{abstract}
\end{frontmatter}

{\it Introduction.} The distinctive characteristic of  Quantum Chromodynamics (QCD) is its ability to dynamically enhance the interaction in the infrared (IR) region or at long distances. Phenomena like the  mass generation of the light hadrons is outside the known Higgs mechanism, but it is born in the enhancement of the IR interaction among quarks and gluons, leading to chiral symmetry breaking and the quark dressing~\cite{Bashir:2012fs,CloPPNP14,horn2016pion,Eichmann:2016yit}. At the same time  gluons are also dressed and in extreme IR region they behave as massive particles of about 600~MeV~\cite{OLJPG11} (for a recent discussion of the gluon propagator in the Landau gauge see~\cite{li2020generalised}). Any successful representation of the QCD dynamics by means of models has
to incorporate the strengthening of the effective interaction in IR region~\cite{Oliveira:2020yac} and dressed degrees of freedom. The interaction among these dressed quarks is based on the exchange of dressed gluons, which has a range of about $\sim 0.3$~fm inside the hadron, where  the interaction at larger distances becomes strong, or below momentum of 600~MeV$\sim 2\Lambda_{QCD}$. Therefore, to some extend  the constituent quarks have a short-range interaction at low momentum, which could be parameterized effectively as a contact one, as happens in the successful phenomenology provided by Nambu-Jona-Lasinio quark models~(see e.g.~\cite{Hatsuda:1994pi}).

The implementation of effective interactions among the constituent quarks on the hypersurface  $(x^+=t+z=0)$ within light-cone quantization~\cite{Brodsky:1997de,Vary:2009gt,BakkerNPB2014,Vary:2016emi}, allows to find the hadron state as an eigenstate of the light-front (LF) Hamiltonian. On the other hand such description  leads to the partonic picture of the hadron in the ultraviolet (UV) momentum region. The hadron state  expanded in the LF Fock basis defined with constituent degrees of freedom, 
ultimately allows to build the hadron image  through the different  probability densities (see e.g.~\cite{Arrington_2021}). 
The associated parton distributions to be studied in future facilities as  the Electron Ion Collider (EIC)~\cite{AbdulKhalek:2021gbh} and the one in China (EicC)~\cite{Anderle:2021wcy}, will provide  precise information on the QCD nonperturbative IR physics.

Among the Fock-components of the hadron wave function, the dominant valence one can also be  obtained from the projection onto the LF of the Minkowski space Bethe-Salpeter  amplitude (see e.g.~\cite{Frederico:2010zh,MezragFBS}). Furthermore,  the valence wave function is an eigenstate of the effective LF mass squared operator reduced to the valence sector, which can be derived from the BS equation using the quasi-potential approach applied to two bosons~\cite{Sales:1999ec}, two fermions~\cite{Sales:2001gk} and to three-particles~\cite{Marinho:2007zz,Guimaraes:2014kor}.  The effective interaction contains the infinite sum over intermediate states in the Fock-space. Alternatively, the eigenvalue equation for the effective LF mass squared operator can be derived using the "iterated resolvent method"~\cite{Brodsky:1997de}. 

In this work, we further study the  proton  within the framework of the LF projected Minkowski space Faddeev-Bethe-Salpeter (LF-FBS) equation for three particles  with  a contact pairwise interaction~\cite{Frederico92,Carbonell03} (see~\cite{Ydrefors:2021mky} for the detailed derivation within the quasi-potential approach). This valence model  has been  used to investigate the  proton structure with the totally symmetric momentum component of the valence wave function~\cite{deAraujo95} and applied recently~\cite{Ydrefors:2021mky} to study the proton image on the null-plane. The essential dynamical ingredient is a diquark~\cite{BarabanoPPNP2021}, which is introduced in the model through the quark-quark amplitude,  analogous to  formulations of the nucleon with Euclidean continuum methods~\cite{Lu:2022cjx}. Other recent approaches to the nucleon emphasize diquark degrees of freedom (see e.g.~\cite{Hobbs:2016xfz,AlvarengaNogueira:2019zcs}). 

 The kernel of the LF-FBS equation applied for the proton is improved  to take into account the separation between the IR and the UV  interaction regions in the three-quark dynamics. This is achieved by  introducing a soft cutoff of the free three-quark LF propagation in states of high virtuality. The deep  unphysical ground state found previously~\cite{Ydrefors:2017nnc} is now naturally removed.   In the present formulation of the LF-FBS model the quark spin degree of freedom is not considered, as it is our goal to study the spatial non-polarized distributions of the quarks in the proton valence state, namely, the parton distribution function (PDF) and transverse momentum distribution (TMD).

 {\it LF three-quark model.} We consider only  the totally symmetric momentum part of the
the colorless three-quark wave function corresponding to the  valence nucleon state, as we are interested  for the time being on the investigation of the properties associated with the momentum distributions and the image of the nucleon onto the null-plane. In this case, the  valence LF wave function is written as~\cite{Ydrefors:2021mky}:
\begin{equation}
\label{Eq:BS_wf}
 \Psi_3(\{x,\vec k_\perp\}) = \sum_{i=1}^3 \frac{ \Gamma(x_i, k_{i\perp})}{\sqrt{x_1 x_2 x_3}\left(M_N^2  
- M^2_0(\{x,\vec k_\perp\})
 \right)},
\end{equation} 
$\{x,\vec k_\perp\}\equiv\{x_1,\vec{k}_{1\perp}, x_2, \vec{k}_{2\perp}, x_3, \vec{k}_{3\perp}\}$ with $\Gamma(x_i, k_{i\perp})$, where $k_{i\perp}=|\vec{k}_{i\perp}|$,  being the Faddeev component of the vertex function for the bound state, $x_1 +x_2 +x_3=1$, 
$\vec{k}_{1\perp}+\vec{k}_{2\perp}+\vec{k}_{3\perp}=\vec 0_\perp$ and
$M^2_0(\{x,\vec k_\perp\}) =\sum_{i=1}^3 \frac{\vec{k}_{i\perp}^2 + m^2}{x_i}\,,$
 is the free three-body squared mass for on-mass-shell constituents.  The factorized form of the valence wave function, namely with a vertex function depending solely on the bachelor quark LF momenta, is a consequence of the effective contact  interaction between the constituent quarks, which is an idealized model resembling the successful Nambu-Jona-Lasinio model applied to model QCD~\cite{Klevansky:1992qe}. It should be understood as an effective low-energy model which is meant to have significance in the IR region where constituent quarks are massive and bound forming the nucleon.  

The bound state homogeneous Faddeev equation for the vertex component of the nucleon valence LF wave function, with four-point local interaction between the constituents
 was described in Refs.~\cite{Frederico92,Carbonell03} and applied to study the proton~in~\cite{deAraujo95,Ydrefors:2021mky}. One physical key ingredient was missing so far in these previous studies, namely the kernel of the dynamical integral equation has to take into account the IR enhancement of the QCD interaction between the quarks~\cite{Oliveira:2020yac} and the weakening at large momentum scale. However, our model four-point local interaction has its action in  the UV region, while in IR  quarks and gluons interact strongly according to QCD. In order to represent the physics of QCD, which undoubtedly distinguish  the IR and UV dynamics, we model the kernel in a way that it  is  
 stronger at low momenta and weaker for large ones.
For this aim we introduced a smooth cutoff in the integral equation for the Faddeev component of the vertex such that:
 \begin{multline}
  \label{Eq:3b_LF_new}
  \Gamma(x, k_\perp) =  \frac{\mathcal{F}(M^2_{12})}{(2\pi)^3}\int_0^{1-x}\frac{dx'}{x'(1-x-x')}\\ \times\int_0^\infty d^2 k'_\perp \frac{\Lambda(\widehat{M}^2_0)}{\widehat{M}_0^2 - M_N^2} \Gamma(x',k'_\perp)\, ,
\end{multline}
where $\mathcal{F}(M^2_{12})$ is the quark-quark amplitude,
$\widehat M_0^ 2=M^2_0(x,\vec{k}_\perp,x',\vec{k}'_\perp, 1-x-x', -(\vec{k}_\perp+\vec{k}'_\perp))\, .$
Eq.~\eqref{Eq:3b_LF_new} for $\Lambda(\widehat{M}^2_0) =1$ was derived in detail in~\cite{Ydrefors:2021mky} resorting to the quasi-potential technique  to perform the projection onto the LF of the  three-boson BS equation in  Minkowski space. Such LF equation corresponds to the truncation of the LF Fock-space only at the valence level. The form factor is introduced to cut the three-quark resolvent at large virtuality, such that
\begin{equation}\label{eq:ffqcd}
      \frac{1}{\widehat{M}_0^2 - M_N^2} - \frac{1}{\widehat{M}^2_0 + \mu^2}=\frac{\Lambda(\widehat{M}^2_0)}{\widehat{M}_0^2 - M_N^2}\, ,
     \end{equation}
    where
$\Lambda(\widehat{M}^2_0)=(M_N^2+\mu^2)/(\widehat{M}^2_0 + \mu^2)\, ,$
which dampens the kernel when $\widehat{M}_0^2 >> \mu^2$, without changing the low momentum region.
In the IR region $F(\widehat{M}^2_0)\sim 1$, as the minimum value of the three-quark free-mass is $3m$,  which is about the nucleon mass. The IR scale of the model is chosen $\mu\sim \Lambda_{QCD}$. Noteworthy that the IR enhancement of the kernel also should simulate the relevance of the coupling of the valence component to the higher Fock-components at large distances.
In addition, the choice of the form factor  eliminates the unphysical solution with $M_N^2<0$ appearing for the bound diquark case when $\Lambda(\widehat{M}^2_0)=1$~\cite{Ydrefors:2021mky}, as we are going to show.

 \begin{table}[]
    \centering
    \begin{tabular}{c c c c c}
     \toprule
  Model &       $m$ [MeV] & $a.m$ & $\mu/m$ & $M_{dq}$ [MeV]  \\
          \midrule
(a) & 366 & 2.70 & 1 & 644 \\
(b)& 362 & 3.60 & $\infty$ & 682 
\\    
(c) & 317 & -1.84 & $\infty$ & - \\
          \bottomrule
    \end{tabular}
    \caption{Model parameters:  constituent quark mass (2nd column),  scattering length in units of $m^{-1}$ (3rd column), cutoff mass in units of $m$ (4th column) and diquark mass (5th column). The nucleon mass is 940~MeV.}
    \label{tab:parameters}
\end{table}

 We observe that  the quark-exchange  kernel of the LF-FBS equation~\eqref{Eq:3b_LF_new} regularized at the mass scale $\mu$ can be recognized by the appearance of the regulated  free resolvent given by Eq.~\eqref{eq:ffqcd}. 
The quark-exchange kernel is also present in the Euclidean four-dimensional three-quark BSE~\cite{Eichmann:2016yit} when diquarks dominate the quark-quark interaction.

The  model takes into account the quark-quark amplitude, $\mathcal{F}(M^2_{12})$, which weights the kernel of the LF-FBS equation for the vertex function, and has the following expression in the limit of an effective contact interaction between the constituent quarks:
\begin{small}
 \begin{equation}
\label{Eq:F_amp}
\mathcal{F}(M^2_{12})=
\frac{\Theta(-M_{12}^2)}{\frac{1}{16\pi^2 y}\log\frac{1+y}{1-y}-\frac{1}{16\pi m a}}
+\frac{\Theta(M_{12}^2)\,\,\Theta(4m^ 2-M_{12}^2)}{\frac{1}{8\pi^2 y'}\arctan y'-\frac{1}{16\pi m a}}\,,
\end{equation}
\end{small}
where the $\Theta(x)$ denotes the Heaviside theta function. and its argument is the effective off-shell mass of the two-quark subsystem squared, given by
\begin{multline}
M^2_{12} = (1-x)M^2_N - \frac{k^2_{\perp}+ (1-x)m^2}{x}\, , 
\,\,\\
y'=\frac{M_{12}}{\sqrt{4m^2 - M^2_{12}}}\, ,\,\,  y=\frac{\sqrt{-M^2_{12}}}{\sqrt{4m^2 - M^2_{12}}}\, . 
\end{multline}

The scalar diquark is a pole of the quark-quark amplitude, Eq.~\eqref{Eq:F_amp},  for scattering lengths $\pi/(2m)>a>0$ associated with a diquark mass $M_{dq}$, which is a model parameter. The  value of $M_{dq}$ is suggested by the recent literature~\cite{BarabanoPPNP2021} to be around 600~MeV.
In the case $a$ is negative no physical two-body bound-state exists and the nucleon is a Borromean state. In both situations, when $\pi/(2m)>a>0$ and  $a<0$, the quark-quark amplitude has a pole. In the former case the pole appears  in the physical complex-energy sheet, while in the latter in the $2^\text{nd}$ one, meaning the virtual state. The strong diquark correlation is manifested either for bound or virtual states and it is a consequence of the enhancement of interaction between the constituent quarks in the  IR region~\cite{BarabanoPPNP2021}. Some choices of parameters of the model   are given in Table~\ref{tab:parameters}, with (a) being the new one for $\mu=m$,  (b) and (c) for $\mu=\infty$, which were already studied  in~\cite{Ydrefors:2021mky}. The choice (a) of model parameters will become clear later on.

\begin{figure}[!t]
    \centering
     \includegraphics[height=5.4cm]{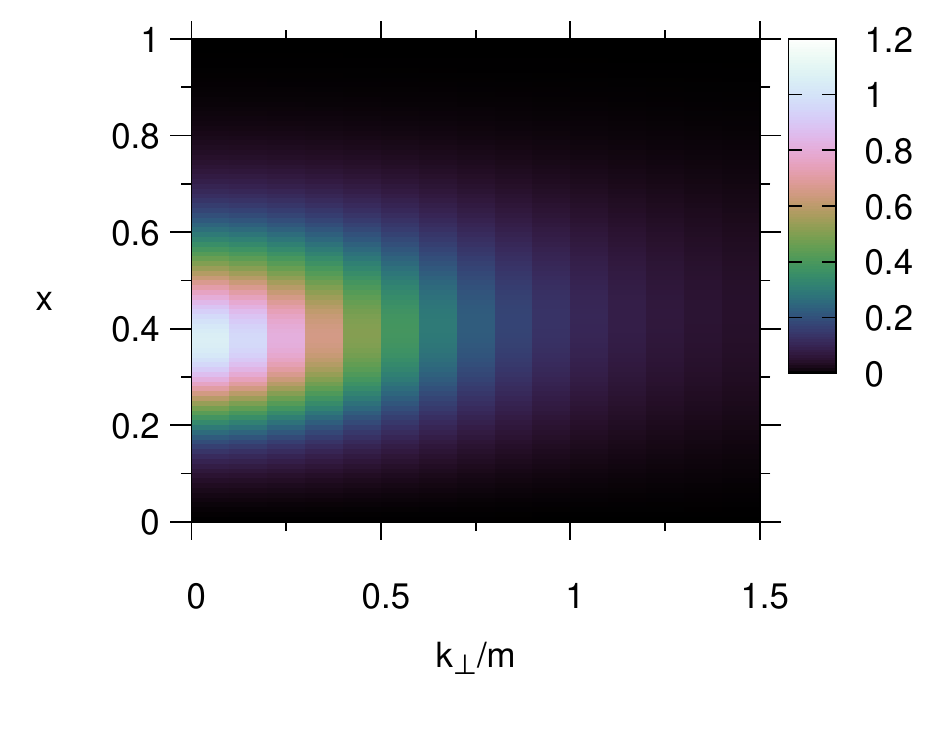}
    \includegraphics[height=4.4cm]{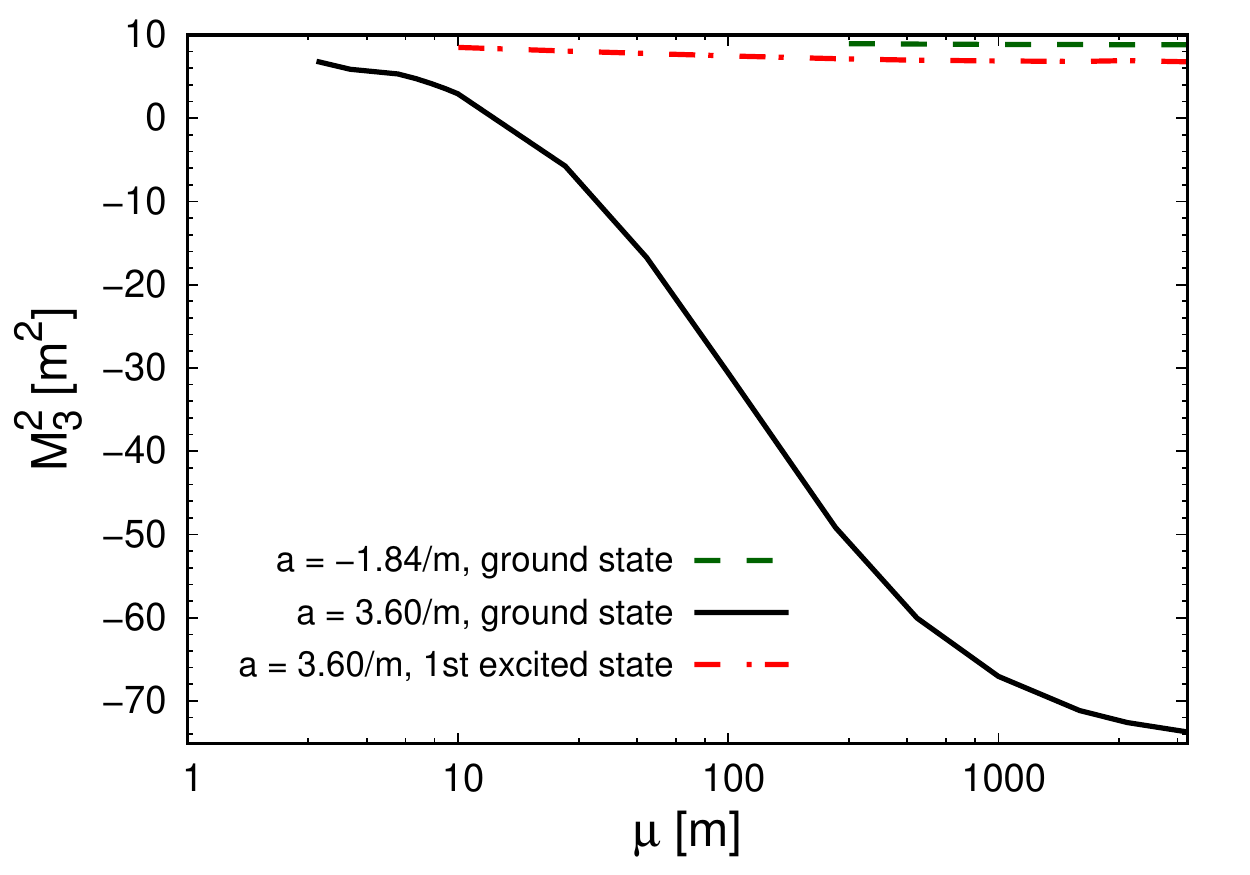}
    \caption{Upper panel: Vertex function, $\Gamma(x,k_\perp)$, in arbitrary units for model (a). 
    Lower panel: Computed values of the squared three-body bound state mass, $M^2_3$, versus the cutoff mass, $\mu$, for $a/m=3.60$ ground  (solid line) and first excited state (dot-dashed line), and  $a/m=-1.84$ (dashed line). The asymptotic limits of $\mu\to\infty$ correspond to model (b) (dot-dashed line) and to model (c) (dashed line).}
    \label{Fig:M_mu}
\end{figure}

In the upper panel of Fig.~\ref{Fig:M_mu}, the Faddeev component of the vertex, $\Gamma(x,k_\perp)$,  is shown. The quark mass in model (a) is chosen close to 350~MeV which is the IR value obtained in a recent LQCD calculation in the Landau gauge~\cite{Oliveira:2018lln}. 
We note that the diquark  mass of $644\,$MeV presents a difference of 278~MeV with respect to the quark mass, comparable  to the gauge invariant result from the LQCD calculation~\cite{Francis:2021vrr} of 319(1)~MeV at the physical pion mass.
For the model (a) with parameters inspired by LQCD results, the three-quark system has only one bound state, which is identified with the nucleon. The vertex function is peaked at $x\sim1/3$ and it spreads out up to $k_\perp\sim \Lambda_{QCD}$, which is about the constituent quark mass and turns even more reasonable our  choice of $\mu$.  In particular, model (a) provides a fit to the Dirac form factor of the proton, as it will be shown.

To be complete, we present in the lower panel of Fig.~\ref{Fig:M_mu}, the computed  values of the squared three-body mass, $M^2_3$, as a function of the cutoff mass $\mu$ for  the ground  and first excited states for  $a=3.6/m$, and ground state in the case of $a=-1.84/m$.  By decreasing the value of $\mu$ the unphysical ground state of the model with $M^2_3<0$ for $a=3.6/m$ disappears and turns to a physical one. A similar effect is found for  $a=2.7/m$ which for $\mu=m$ defines the model parametrization (a) given in Table~\ref{tab:parameters}.

\begin{figure}[!t]
  \centering
\includegraphics[height=4.4cm]{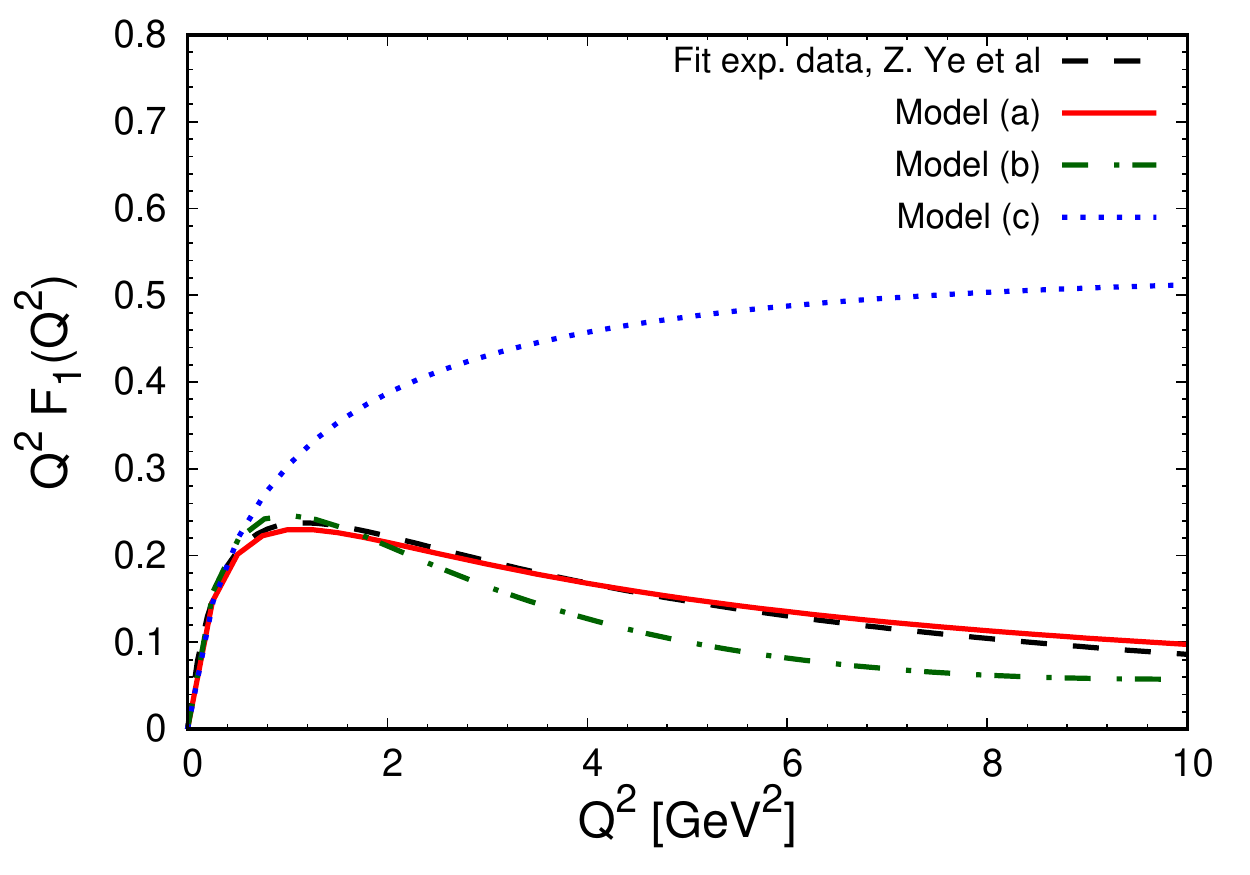} 
  \caption{ 
 $F_1(Q^2)$ for 
 model (a) (solid line), 
  (b) (dot-dashed line) and 
 (c) (dotted line).
 The empirical fit (dashed line) obtained in Ref.~\cite{Ye18}.
  \label{Fig:FF}}
\end{figure}

{\it  Dirac form factor.} The present model accounts only for the valence state of the nucleon wave function, and it allows to obtain  the Dirac  form factor as  discussed in detail in~\cite{Ydrefors:2021mky}. Resorting to the Drell-Yan condition where the plus component of the momentum transfer vanishes $(q^+=0)$, the form factor computed with the valence component  of the wave function is given by:
\begin{equation}
  \label{Eq:F_1}
  F_{\text{1}}(Q^2) =  \int\{dx\,d^2k_\perp\}\,
  \Psi_3^\dagger(\{x,\vec{k}^\text{f}_{\perp}\})\Psi_3(\{x,\vec{k}^\text{i}_{\perp}\}),
\end{equation}
where the phase-space integral is
\begin{equation}
 \int\{dx\,d^2k_\perp\}=    \prod_{i=1 }^2\int\frac{d^2k_{i\perp}}{(2\pi)^2} \int^1_0\hspace{-.2cm} dx_i\,\Theta\big(1-x_1-x_2\big)\,,
\end{equation}
with $\sum_{i=1}^3\vec k_{i\perp}=0$, $\sum_{i=1}^3x_i=1$ and $Q^2 = \vec{q}_\perp \cdot \vec{q}_\perp$. Furthermore,  choosing the Breit frame the momenta of the quarks in Eq.~\eqref{Eq:F_1} are:
\begin{multline}
\label{Eq:defs0}
 \vec k^\text{f(i)}_{i\perp}=\vec{k}_{i\perp} \pm \frac{\vec{q}_\perp}{2}x_i\,\,\,(i=1,2) \quad\text{and}\quad \\
 \vec k^\text{f(i)}_{3\perp}=   \pm \frac{\vec{q}_\perp}{2}
 (x_3-1) - \vec{k}_{1\perp} - \vec{k}_{2\perp}\, ,  
\end{multline}
with -(+) for f(i).

In  Fig.~\ref{Fig:FF}, we present the results for the proton Dirac form factor, $F_1(Q^2)$, with model (a), which was fine tuned to be consistent with the global fit to  experimental data by Ye et al \cite{Ye18} up to 10~GeV$^2$. In addition, we present the previous results~\cite{Ydrefors:2021mky} obtained with model (b) having a bound diquark, and with model (c) presenting a virtual diquark, both calculations have  $\mu=\infty$. The IR  dynamics   is not privileged by models (b) and (c) and the  calculated  Dirac form factor is not able to  reproduce the experimental fit, which is now possible. The enhancement of the interaction kernel of  Eq.~\eqref{Eq:3b_LF_new} in the IR  with respect to the  UV region,  was achieved with the introduction of the regularization scale $\mu\sim\Lambda_{QCD}$ in Eq.~\eqref{eq:ffqcd}, which finally lead the model to the experimental  Dirac form factor.

\begin{figure}[t]  \centering  \includegraphics[height=4.4cm]{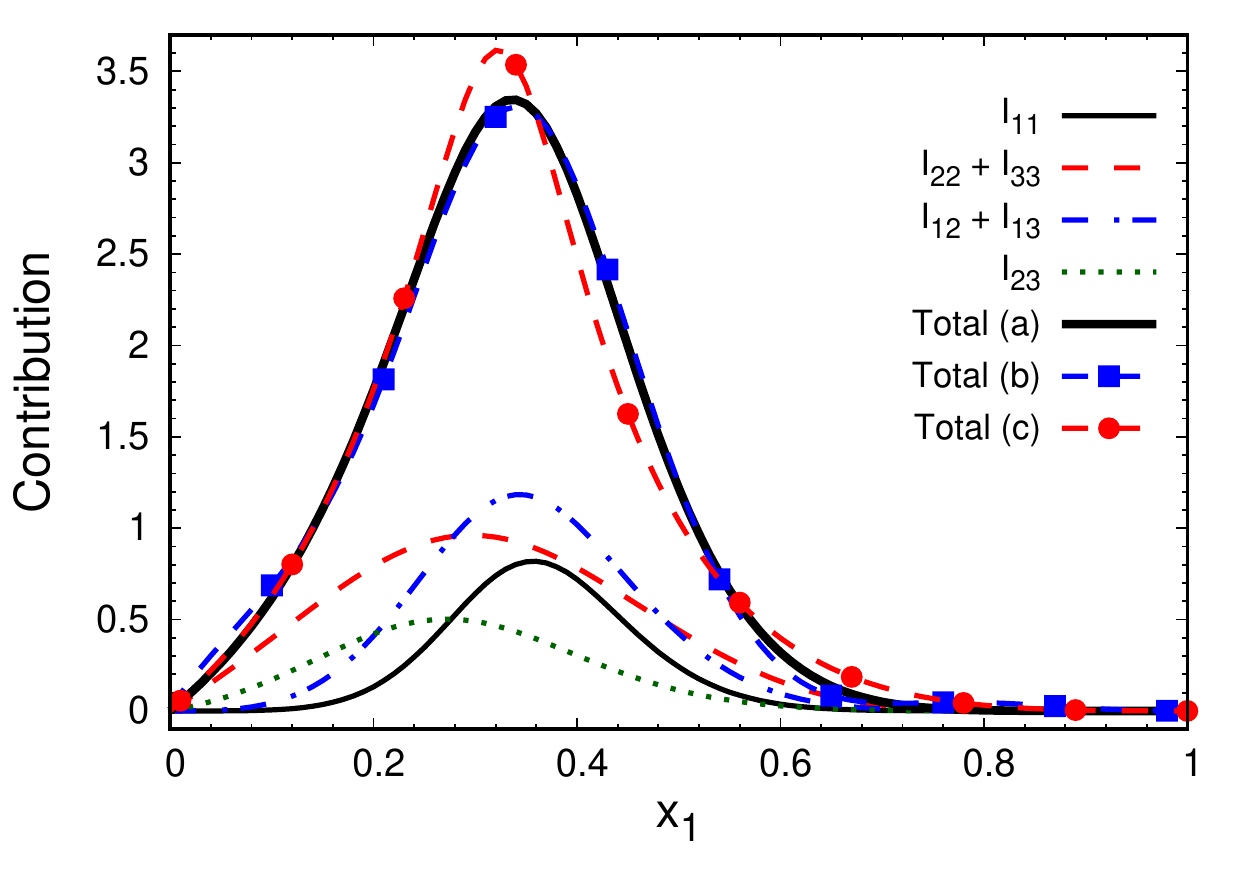} 
\caption{ Valence PDF computed with model (a): $I_{11}$ (solid thin line), $I_{22}+I_{33}$ (dashed line), $I_{12}+I_{13}$ (dot-dashed line), $I_{23}$ (dotted line) and the total (solid thick line) normalized to 1. Results for the total PDF for model (b) (full squares) and (c) (full circles) are connected by the dashed lines. 
\label{Fig:PDFcont}}\end{figure}

{\it Valence quark momentum distribution.} 
The valence PDF  should be given by  the sum over all Fock components of the proton light-front wave function, namely:
\begin{equation}
\label{Fockpdf}
\begin{aligned}
    f_1(x)=& 
    \sum_{n=3}^\infty   \,
    \Bigg\{ \prod_{i}^n \int\frac{d^2k_{i\perp}}{(2\pi)^2} \int^1_0 dx_i\Bigg\}\, \\
    &\times \delta\left(x-x_1\right)\delta\left(1-\sum_{i=1}^n x_i\right)\, \delta\left(\sum_{i=1}^n\vec k_{i\perp}\right) \\
    &\times \big|\Psi_n(x_1,\vec k_{1\perp},x_2,\vec k_{2\perp},...)\big|^2\, ,
\end{aligned}
\end{equation}
where $n$ indicates the number of partons in each Fock contribution to  the probability amplitude. However, in   the present model, we consider only the valence component  of the proton LF wave function, namely  the truncation takes into account  only $n=3$, which presumably  is  the  dominant one. It has been verified for the  pion state~\cite{dePaula:2020qna}  where the valence contribution accounts for 70\%  of the LF wave function described in terms of   constituent  quark degrees  of freedom.

The results for the valence parton distribution (PDF) at the hadron scale are shown in  Fig.~\ref{Fig:PDFcont} for model (a), (b) and (c). The PDF is  obtained from the integrand of  Eq.~\eqref{Eq:F_1} for  the Dirac form factor at $Q^2=0$  or equivalently from Eq.~\eqref{Fockpdf} truncated at the valence state:
 \begin{equation}
  \label{Eq:pdf_Q20}
  {f}_1(x) = \sum_{3\ge j\ge i\ge 1} 
   (2-\delta_{ij}) \, I_{ij}(x)  \, ,
\end{equation}
where
\begin{small}
\begin{equation}\label{Eq:pdf_Iij}
 I_{ij}(x)=   \int \{dx\,d^2 k_\perp\}
   \frac{\delta(x-x_1)}{x_1x_2x_3}
   \frac{\Gamma(x_i, \vec{k}_{i\perp})\Gamma(x_j, \vec{k}_{j\perp})}{\big(M_N^2 - M^2_0(\{x,\vec{k}_{\perp}\})\big)^2}\, ,
\end{equation}
\end{small}
with  $\sum_{i=1}^3\vec k_{i\perp}=0$ and $\sum_{i=1}^3x_i=1$.
The contributions to the PDF indicated in the figure are identified by  $I_{ij}$ defined in Eq.~\eqref{Eq:pdf_Iij}. Due to the symmetry of the  nucleon wave function under the exchange of quarks 2 and 3, it follows that $I_{22} = I_{33}$ and $I_{12} = I_{13}$, these relations are  taken into  account in Eq.~\eqref{Eq:pdf_Q20}.
Noteworthy  to find  that all the contributions have about the same size, and are peaked around 1/3, despite our choice of quark 1 to obtain  the PDF. This property can be traced back to the denominator appearing in the  valence wave function in Eq.~\eqref{Eq:BS_wf}, which is the three-quark resolvent and has its  maximum value at the smallest virtuality of the three-quark system. The contribution from $I_{11}$ corresponds to the situation where the quark 1 is picked up while the pair of quarks 2 and 3 are in a diquark correlation, which is a small fraction of the total PDF, showing that the symmetrization of  the valence wave function is relevant for building the  proton PDF. 

All contributions to the PDF are indeed similar in magnitude, and no one is dominant, as we have already shown previously in the study of models (b) and (c) in Ref.~\cite{Ydrefors:2021mky}. Interesting to observe that models (a) and (b), which both present a bound diquark and provide results close to the experimental  proton Dirac form factor, as seen in Fig.~\ref{Fig:FF}, also have very  similar PDF's. This suggests that for these observables, form factor and PDF, the formation of the diquark is a dominant feature, quite independent on the cutoff mass. However,  model (b) presents an unphysical ground state, which is eliminated by the introduction of the cutoff, as we have discussed.

\begin{figure}[!t]
\centering
\includegraphics[height=4.4cm]{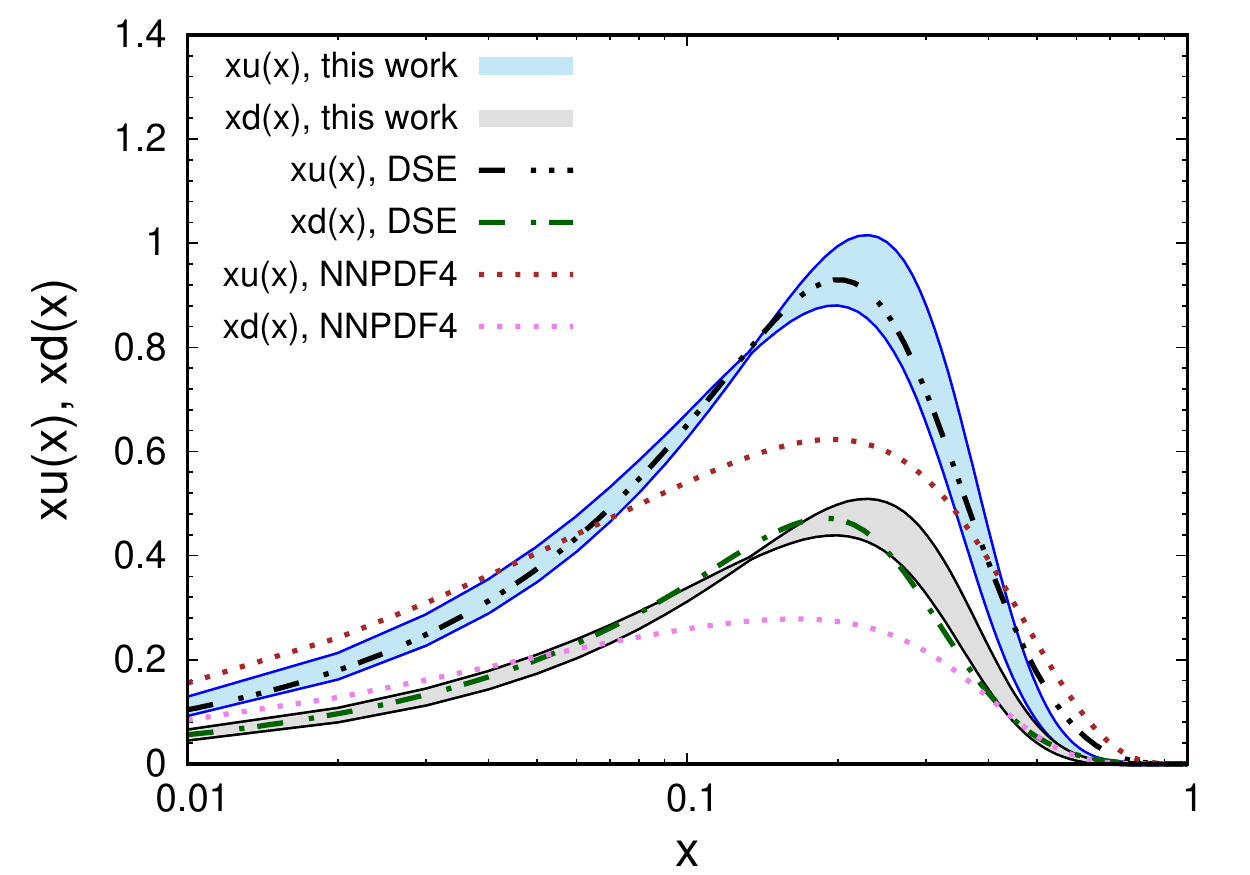}
\caption{
Valence $u$ and $d$ quark PDFs evolved to $Q=3.097\,$GeV. Comparison with   DSE~\cite{Lu:2022cjx}. The shaded areas correspond to model (a) for 
$Q_0=0.33 \pm 0.03\,$GeV.  NNPDF 4.0 global fit~\cite{NNPDF:2021njg} (dotted lines).}
\label{Fig:PDF_evolved}
\end{figure}

{\it Evolved valence PDFs.} We present results for model (a) at $Q=3.097\,$GeV for the $u$ and $d$ valence quarks in Fig.~\ref{Fig:PDF_evolved}. We compare 
 with the recent results obtained with the Dyson-Schwinger Equation (DSE) approach~\cite{Lu:2022cjx}. In this work we adopted the same method as the one used in~\cite{dePaula:2022PRDL} for the evolution of the pion PDF. Namely, the DGLAP equation with lowest-order splitting function was used together with the effective charge from~\cite{Cui2022}.

We also took into account an uncertainty with respect to the initial proton scale: $Q_0=0.33 \pm 0.03\,$GeV~\cite{Cui2022}, and compared our calculations with the results from the NNPDF 4.0 global fit\footnote{Shown in the Fig.~\ref{Fig:PDF_evolved} are the central values of the NNPDF~4.0 global fit available in the PDF database at \url{https://lhapdf.hepforge.org/pdfsets.html} and the data was extracted by using the LHAPDF software \cite{LHAPDF}.}~\cite{NNPDF:2021njg}. The model (a) produces  results quite consistent with the DSE approach, which privileges the strong diquark correlation, only at large $x$ the discrepancy becomes visible with a softer behavior from the DSE approach. The present model has a strong UV damping due to the cutoff, being not asymptotically free at large momentum, which is associated with a harder behavior at the end-point. Furthermore, it  has only  the  contribution  for the PDF  from the valence light-front wave function, which is strongly  peaked around  1/3. It is expected  that the higher Fock-components contribution to  the PDF moves the  peak towards smaller $x$, as the  proton longitudinal momentum is shared between more than three quarks. Therefore,   the value of $\langle x_q \rangle$ should be somewhat decreased, as we observe from the comparison of model (a) with NNPDF4.0 in Table~\ref{tab:moments} for the first Mellin moments. 

In the present  LF-FBS model the proton has  100\% probability to be in the valence state, as we have considered only  $n=3$ in the Fock space decomposition of Eq.~\eqref{Fockpdf}. We verified that the sum rule $\langle x_u\rangle+\langle x_d\rangle=1$ is saturated at the proton initial scale by adopting the standard normalization:
\begin{equation}
    \int_0^1 dx\, f_u(x)=2 \quad \text{and} \quad \int_0^1 dx \, f_d(x)=1\, ,
\end{equation}
where $f_q(x)=(2-\delta_{q,d})f_1(x)$  with $f_1(x)$ normalized to 1 and obtained
from Eq.~\eqref{Eq:pdf_Q20}.  
The assumption of only valence state in the proton at the initial scale is not fully sustained by the comparison with NNPDF4.0 global  fit as shown in Table~\ref{tab:moments} at $Q=3.097\,$GeV, as the model Mellin moments  $\langle x_u\rangle$ and $\langle x_d\rangle$ are clearly over estimated.
On  the other hand the  comparison with the continuum DSE results from Ref.~\cite{Lu:2022cjx}, which takes into account the detailed spin structure of the quarks and diquarks, suggests  that the  spin contribution is averaged out in the non-polarized PDF,
as  shown in both Fig.~\ref{Fig:PDF_evolved} and in Table~\ref{tab:moments} for the first few moments at  $Q=3.097\,$GeV. 

Noteworthy, the present model represents the truncation of the FBS  equation at  the valence level, and its full  representation on the LF has to  consider the contribution of an induced  three-body interaction  from  the coupling of the valence  with higher Fock-states~\cite{Karmanov:2009bhn,Ydrefors:2017nnc}. In principle,  this four-dimensional FB equation model  only takes into account the quark sea, and despite of that a quite relevant effect in the binding energy was found in~\cite{Ydrefors:2017nnc}. Here, with the introduction of the soft cutoff, this effect would be somewhat reduced. However, gluons are not taken into account in this model which presumably are the main degrees of freedom to share  the quark longitudinal momentum.

\begin{table}[!tp]
    \centering
    \begin{tabular}{c c c c c}
    \toprule
 $q$   & $u$ & $d$ \\ 
     \midrule
    Model  (a) & 
   \\
    $\langle x_q \rangle$  & $0.296 \pm 0.025$ & $0.148 \pm 0.012$ \\
    $\langle x_q^2 \rangle$  & $0.071 \pm 0.009$ & $0.036 \pm 0.005$ \\
    $\langle x_q^3 \rangle$ & $0.021 \pm 0.004$ & $0.011 \pm 0.002$ \\
    $\langle x_q^4 \rangle$  & $0.007 \pm 0.002$ & $0.004 \pm 0.001$ \\
 DSE~\cite{Lu:2022cjx}        & 
   \\
             $\langle x_q \rangle$  & $0.303$ & $0.137$ \\
    $\langle x_q^2 \rangle$  & $0.077$ & $0.032$ \\
    $\langle x_q^3 \rangle$  & $0.032$ & $0.009$ \\
   $\langle x_q^4 \rangle$  & $0.010$ & $0.003$ \\
  NNPDF4.0 & \\
    $\langle x_q \rangle$  & $0.261$ & $0.101$ \\
    $\langle x_q^2 \rangle$  & $0.072$ & $0.023$ \\
    $\langle x_q^3 \rangle$  & $0.027$ & $0.007$ \\
    $\langle x_q^4 \rangle$  & $0.012$ & $0.003$ \\
    \bottomrule 
    \end{tabular}
    \caption{Mellin moments, $\langle x^n_q\rangle$,  of the valence quark PDF $(q=u,d)$ at  $Q=3.097\,$GeV for model (a) compared with the DSE~\cite{Lu:2022cjx} and 
    the NNPDF 4.0 global fit~\cite{NNPDF:2021njg}, obtained by integrating the PDFs shown in Fig.~\ref{Fig:PDF_evolved}.}
    \label{tab:moments}
\end{table}

{\it Transverse  momentum distribution at the proton scale.} The single quark transverse  momentum distribution 
 in the forward limit~\cite{Lorce_2011} 
is
associated with the probability density to find a quark with momentum $k_{\perp}$ and  $x$, when  truncated to  the valence  component is:
\begin{small}
\begin{equation}
\label{eq:TMD}
\begin{aligned}
    \tilde{f}_1(k_\perp,x) &=
      \int_0^1dx_1\hspace{-0.0cm}\delta(x-x_1)\int \frac{dk_{1\perp}}{(2\pi)^2}\delta(k_\perp - k_{1\perp} )\\
    &\times \int_0^{2\pi}d\theta_1\int \frac{d^2 k_{2\perp}}{(2\pi)^2}\hspace{-0.0cm} \int_0^{1-x} dx_2\, |\Psi_3(\{x,\vec{k}_{\perp}\})|^2\, ,
\end{aligned}
\end{equation}
\end{small}
where only the dependence  on $k_\perp=|\vec{k}_\perp|$ remains due to the  symmetry of the wave function under  rotations in the transverse plane.

The  PDF is the  integrated  TMD on the transverse momentum:
\begin{equation}
    f_1(x)=\int dk_\perp k_\perp\tilde{f}_1(k_\perp,x)
\end{equation}
and the integrated TMD in the longitudinal momentum is
\begin{equation}
 \label{TMDq}
 L_1(k_{\perp}) = k_{\perp}\int_0^1 dx\tilde{f}(k_{\perp},x)
 \, ,
 \end{equation} 
 which represents  the probability  density  of a single quark with transverse momentum $k_{\perp}$.
 
\begin{figure}[!t]
  \centering
  \includegraphics[height=5.4cm]{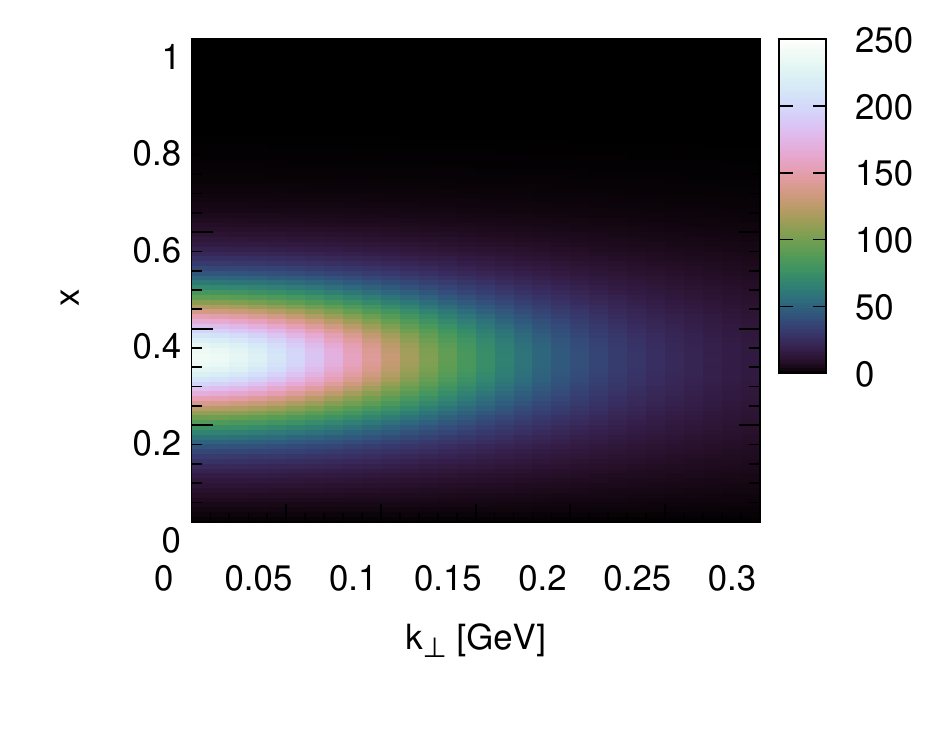}
   \includegraphics[height=4.4cm]{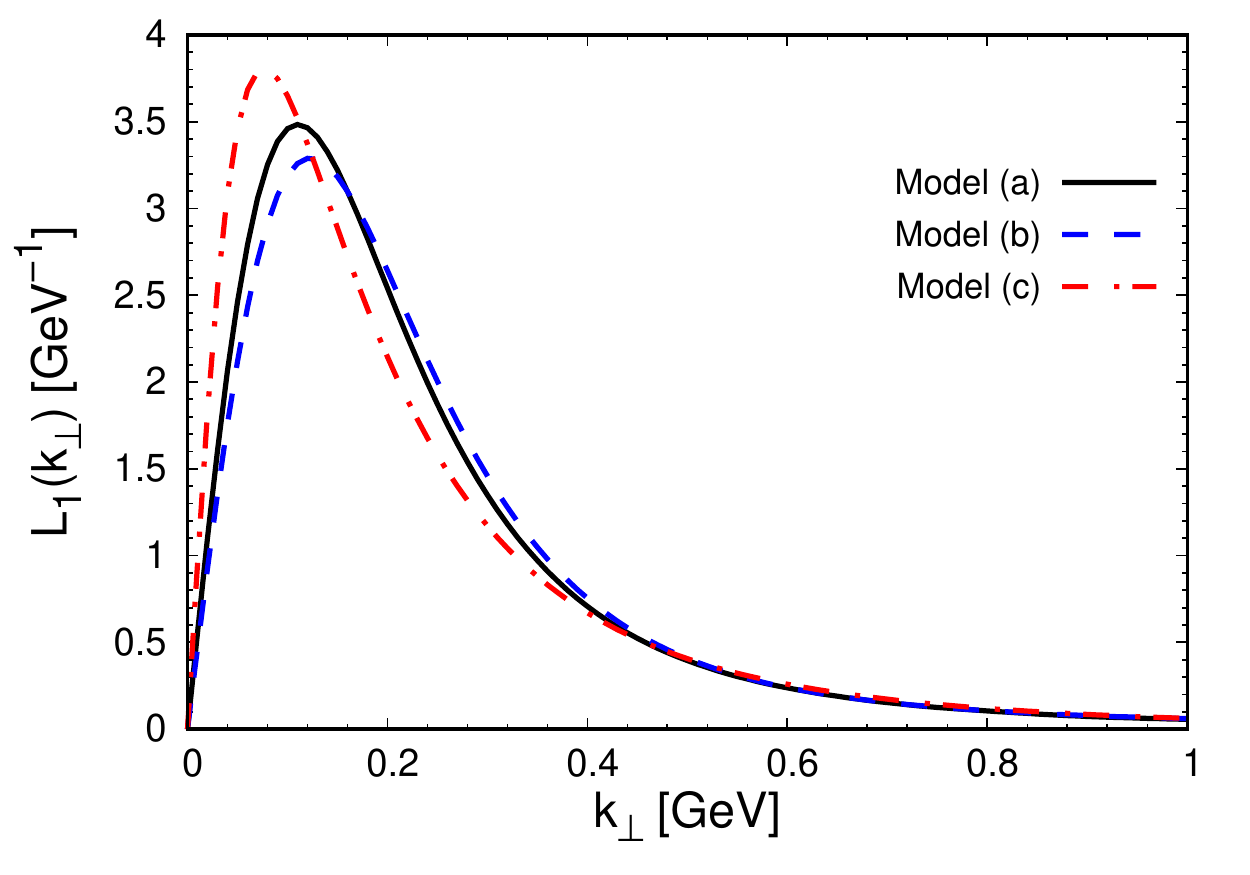}
 \vspace{-0.3cm}
  \caption{Transverse momentum distribution at the proton scale.  Upper panel:
  transverse momentum distribution, 
  $\tilde{f}_1(k_{\perp},x)$ in GeV$^{-2}$  for model (a). Lower panel:  integrated  transverse  momentum density vs $k_{\perp}$ for  the  (a), (b) and (c) models.
  \label{Fig:tmd_1q}}
\end{figure}

In the upper panel of Fig.~\ref{Fig:tmd_1q} the  valence TMD given by Eq.~\eqref{eq:TMD} is presented for model (a), and in the lower panel the result for the single quark transverse momentum  density from Eq.~\eqref{TMDq},  for models (a),  (b) and (c) are shown.  It is interesting to notice that the momentum scale that dominates the TMD and the integrated one  is about $0.15\,$GeV, deep in the IR region, which is associated with the proton size in the transverse direction of $\sim 1.3\,$fm.  As seen in the lower panel of the figure, the models (a) and (b) with a bound diquark present quite similar result, and model (c) with the virtual diquark peaks at considerably lower transverse momentum, reflecting the lower binding energy (see Table~\ref{tab:parameters}).  The TMD and the  integrated one reflect and are a source of information on the IR dynamics of QCD, which in part is reflected in the binding energy of the present model.

{\it Summary.} We further develop the LF effective three-quark model for the proton, based on the notion of the  dynamics dominated by the formation of diquarks. We have disregarded the  momentum structure of quark-diquark vertex function, where the quark-quark amplitude is  obtained from a contact interaction.   The improved version of the proton model includes a soft cutoff  in the kernel of the LF Faddeev equation for the vertex function~\cite{Ydrefors:2021mky}.  This  soft cutoff allows to separate the IR and UV interaction regions, with the physical effect of damping  the contributions from  three-quark configurations at large virtualities. This new development eliminated the unphysical ground state for bound diquarks achieved in previous calculations~\cite{Ydrefors:2017nnc}.
The present model was tuned to reproduce the Dirac proton form factor with a reasonable set of parameters: a quark mass of $366\,$MeV, a diquark mass of $644\,$MeV and a cutoff of $366\,$MeV $(\sim \Lambda_{QCD})$, which was enough  to give the proton mass. From this parameter set we explored the proton non-polarized  quark longitudinal and transverse momentum distributions obtained from the valence wave function. We found that the explicit consideration of the spin degree of freedom of both quark and  diquark is not relevant for the evolved non-polarized valence PDF. However, the comparison  with the global  fit from  the NNPDF~4.0 at $Q=3.097\,$GeV suggests that the higher Fock-components, missed in the model wave function, could be relevant to improve the valence PDF.

Future challenges for the advance of the present nucleon effective model: the  treatment Bethe-Salpeter amplitude in the four-dimensional Minkowski space~\cite{Ydrefors:2019jvu,Ydrefors:2020duk},  the spin degree of freedom for polarized PDFs, and  quark dressing, which will lead to further insights into the nucleon structure.
 
{\it
This work is a part of the project INCT-FNA \#464898/2014-5.
This study was financed in part by Conselho  Nacional de Desenvolvimento Cient\'{i}fico e  Tecnol\'{o}gico (CNPq) under the grant 308486/2015-3 (TF). E.Y. thanks for the financial
support of the grants \#2016/25143-7 and \#2018/21758-2 from FAPESP.  We  thank the FAPESP Thematic  grants    \#2017/05660-0 and \#2019/07767-1.  }

\end{document}